\documentclass[twocolumn]{aastex631}
\usepackage{amsmath}
\usepackage{physics}

\usepackage{pythonhighlight}

\usepackage{xcolor}  

\hypersetup{
  colorlinks=true,
  linkcolor=blue,
  citecolor=blue,
  filecolor=green,
  urlcolor=blue
}

\begin{document}

\title{[PK2008] HalphaJ115927 and IGR J14091-6108$:$ Two new intermediate polars above the period gap}

\email{Email: ajoshi@astro.puc.cl, aartijoshiphysics@gmail.com}

\author[0000-0001-9275-0287]{Arti Joshi}
\affiliation{Institute of Astrophysics, Pontificia Universidad Católica de Chile, Av. Vicuña MacKenna 4860, 7820436, Santiago, Chile}

\begin{abstract}
This study presents a detailed timing analyses of two cataclysmic variables (CVs), [PK2008] HalphaJ115927 and IGR J14091-610, utilizing the optical data from the Transiting Exoplanet Survey Satellite (TESS). Periods of 7.20$\pm$0.02 h, 1161.49$\pm$0.14 s, and 1215.99$\pm$0.15 s are presented for [PK2008] HalphaJ115927, and are interpreted as the probable orbital, spin, and beat periods of the system, respectively. The presence of multiple periodic variations suggests that it likely belongs to the intermediate polar (IP) category of magnetic CVs. Interestingly, [PK2008] HalphaJ115927 exhibits a unique and strong periodic modulation at 5.66$\pm$0.29 d, which may result from the precession of an accretion disc, similar to the IP TV Col. The detection of a spin signal of 576.63$\pm$0.03 s and inferred orbital signal of $\sim$ 15.84 h supports the classification of IGR J14091-610 as an IP. The identification of such a long orbital period adds a new example to the limited population of long-period IPs. The observed dominant signal at the second harmonic of the orbital frequency also suggests ellipsoidal modulation of the secondary in this system.  The observed double-peaked spin pulse profile in [PK2008] HalphaJ115927 likely results from two-pole accretion, where both poles contribute to the spin modulation, and their geometry allows equal visibility of both accreting poles. In contrast, IGR J14091-610 exhibits a single-peaked sinusoidal like spin pulse, attributed to the changing visibility of the accretion curtains due to a relatively low dipole inclination. The present observations indicate that accretion in both systems occurs predominantly through a disc.
\end{abstract}

\keywords{}


\section{Introduction}
\label{sec:intro}
Magnetic Cataclysmic Variables (CVs) are semi-detached binary systems in which a primary magnetic white dwarf (WD) accretes material from a Roche-lobe-filling late-type secondary star \citep{Warner95}. In the intermediate polar (IP) subclass of magnetic CVs, the WD has a magnetic field typically less than $10^7$ G, which is not strong enough to completely disrupt the accretion disc as it does in the polar subclass, where stronger fields prevent any disc formation. In IPs, a partial disc forms but is truncated when the magnetic pressure exceeds the ram pressure, channeling material along magnetic field lines instead. The matter in the accretion column impacts the poles of the WD with supersonic velocities, creating strong shocks that produce cyclotron radiation in the optical/infrared and thermal bremsstrahlung in the X-rays. The optical and X-ray signals in IPs are modulated on the spin ($\omega$), orbital ($\Omega$), beat ($\omega-\Omega$) frequencies, as well as other sidebands  \citep{Patterson94}. A key feature of these systems are the presence of multiple periodicities in the X-ray and optical bands, arising from interactions between spin and orbital modulations either through variations of the accretion rate or through the reprocessing of primary radiation by other components of the system. The existence of multiple periodic components is a key diagnostic tool for identifying the IP nature of candidates in this class. In IPs, accretion can occur via three primary mechanisms: disc-fed, disc-less (or stream-fed), and disc-overflow. These modes of accretions can be identified with the presence of a wide range of X-ray and optical frequencies \citep{Rosen88, Hameury86, Wynn92, Hellier93}. This paper presents detailed optical analyses of two CVs, [PK2008] HalphaJ115927 and IGR J14091-6108. A summary of the available information on these sources is provided below.

\begin{table*}
\small
\centering
\caption{Periods corresponding to dominant peaks in the LS power spectra of H115927 and IGRJ14091.}\label{tab:ps}
\setlength{\tabcolsep}{0.08in}
\begin{tabular}{lcccccccccc}
\hline\\
Object  && Sector & \multicolumn{6}{c}{Periods} \\
\cline{4-9}
&&        &  $P_{\Omega}$  & $P_{2\Omega}$ & $P_\omega$$_-$$_{\Omega}$ &$P_{\omega}$ & $P_{2\omega}$  & $P_{3\omega}$\\ 
&&        &  (h)     &  (h)  &(s)           & ( s )                            & (s)            & ( s )        \\
\hline\\
H115927     && 64           & $7.20\pm0.02$ & $-$  &$1215.99\pm0.15$           &$1161.49\pm0.14$             &$580.74\pm0.04$             & $387.16\pm0.02$ \\
IGRJ14091 && 65   & $-$ & 7.92$\pm$0.02 & $-$ & 576.63$\pm$0.03  & $-$ & $-$ \\
\hline
\end{tabular}
\end{table*}

\begin{figure*}
\centering
\includegraphics[width=0.68\textwidth]{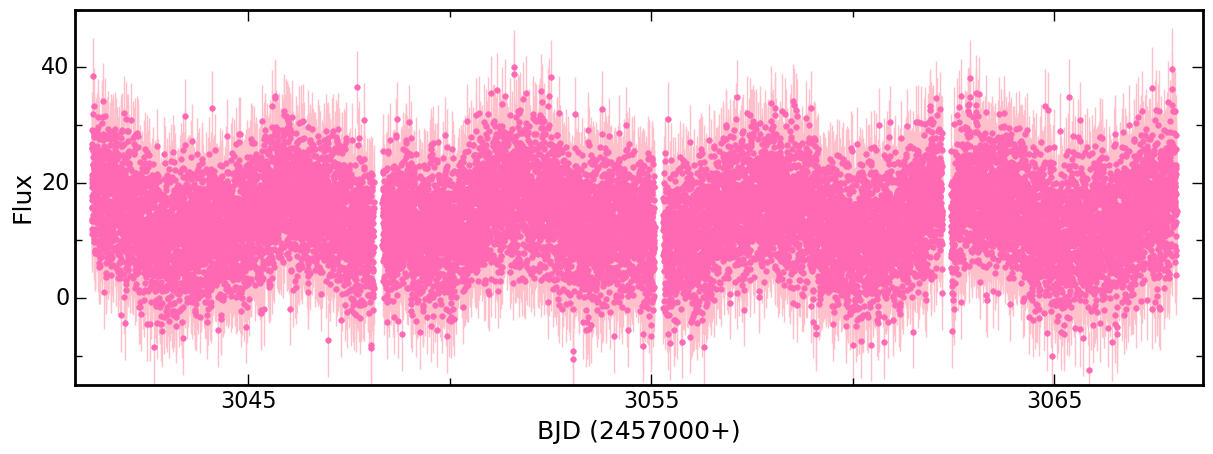}
\caption{TESS light curve of H115927.} 
\label{fig:tesslc_H115927}
\end{figure*}

\begin{figure*}
\centering
\includegraphics[width=0.68\textwidth]{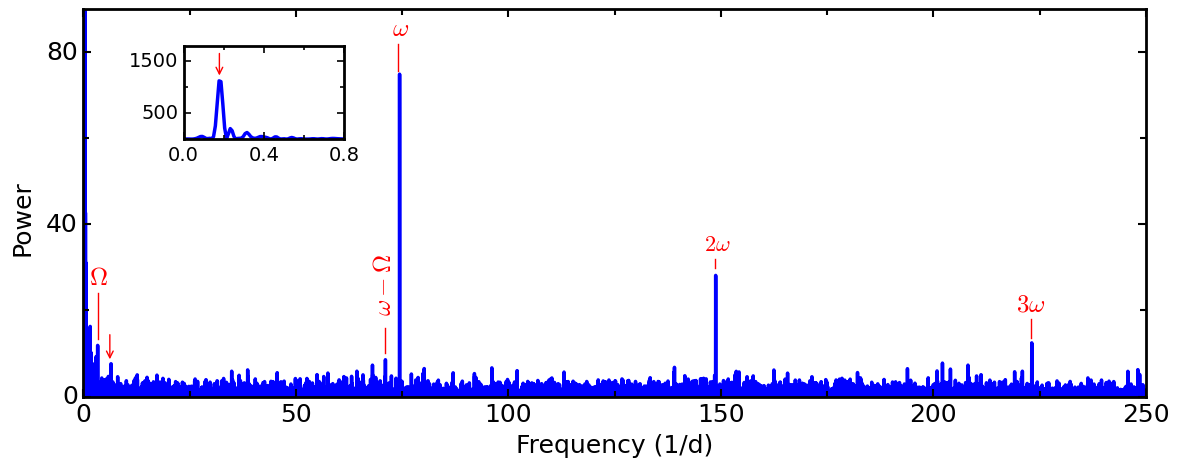}
\caption{TESS power spectrum of H115927, with the inset panel representing a zoomed-in plot of the lower frequency region. The significant frequencies observed in the power spectrum are marked.} 
\label{fig:tessps_H115927}
\end{figure*}

\begin{figure}
\centering
\includegraphics[width=0.45\textwidth]{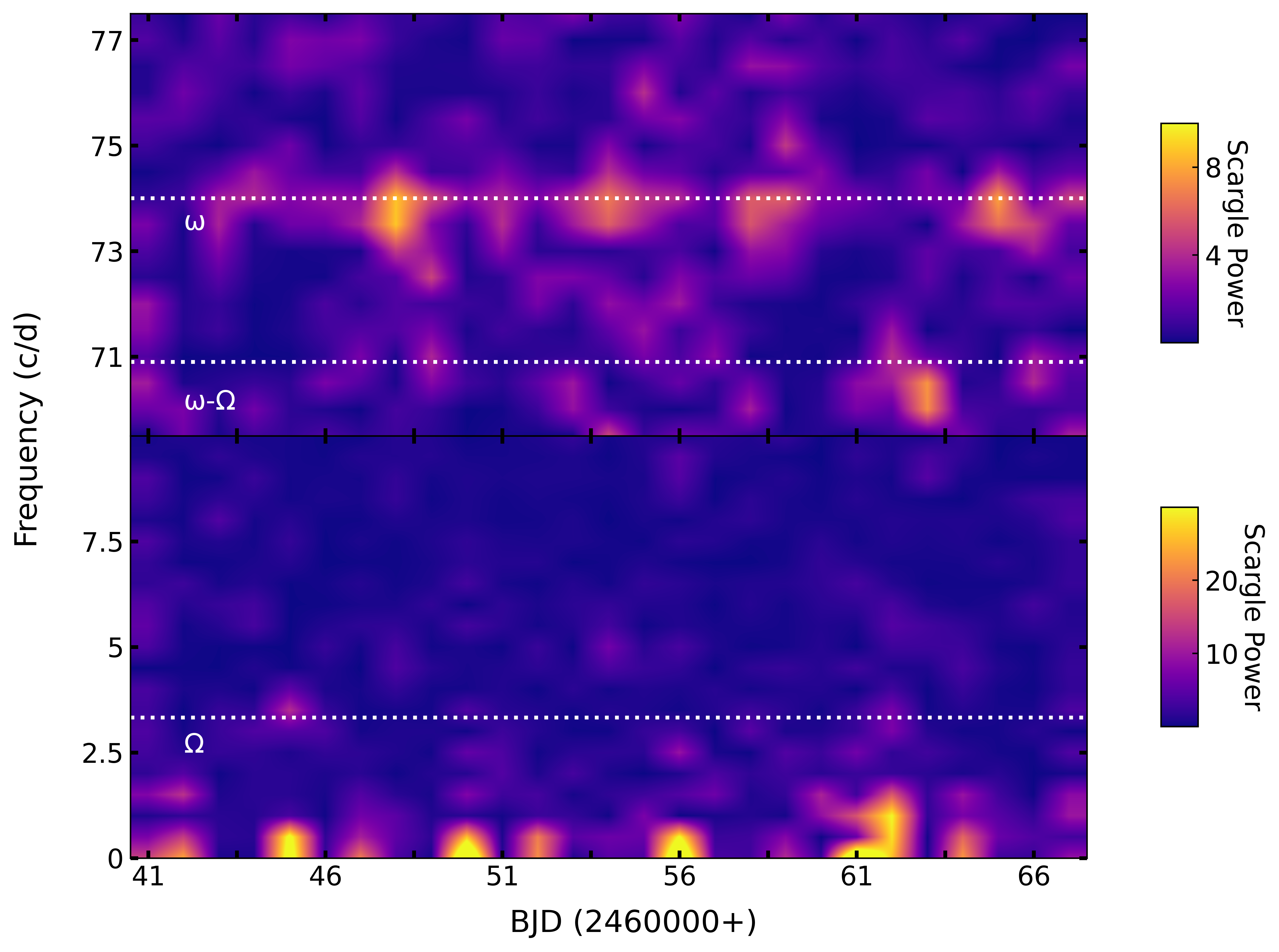}
\caption{Time-resolved power spectra of H115927 computed using a 1-day wide moving window across the entire dataset of the TESS observations near $\Omega$, $\omega-\Omega$, and $\omega$ frequency regions.} 
\label{fig:tessps_1day_H115927}
\end{figure}

\begin{figure}
\centering
\includegraphics[width=0.45\textwidth]{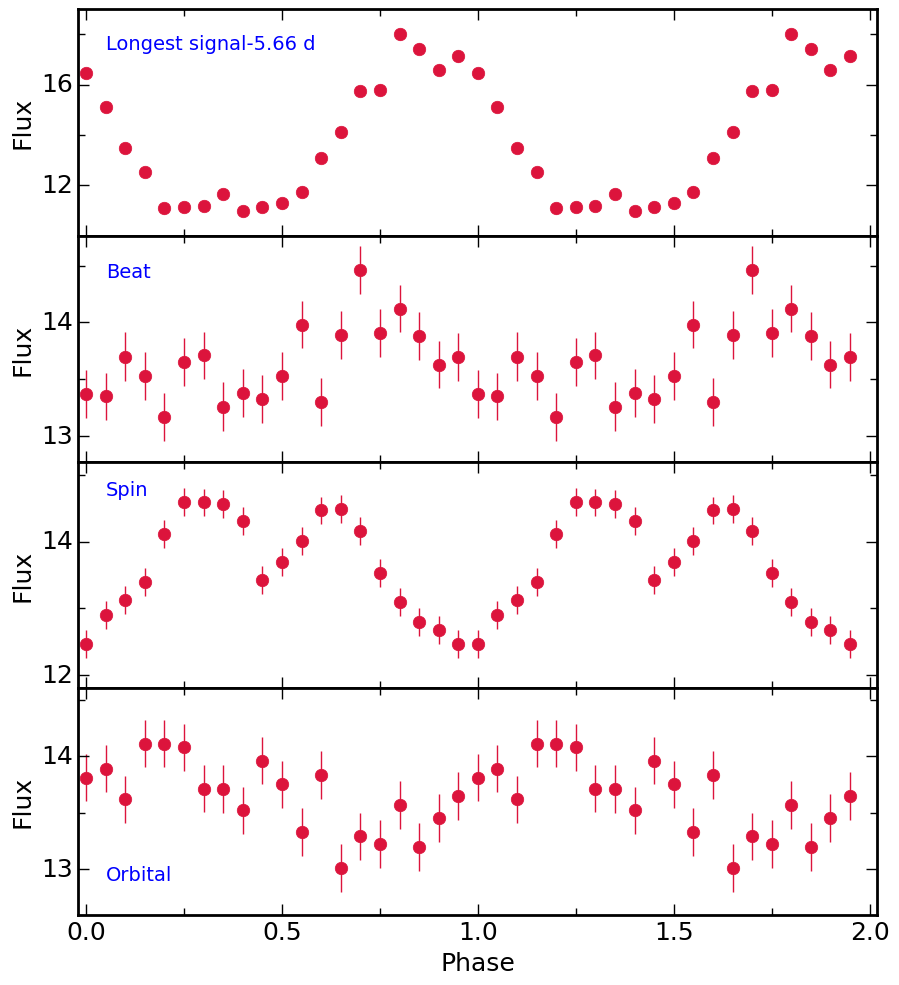}
\caption{Top to bottom panels exhibit the phase-folded light curve variations at the longest signal (5.66 d), followed by the beat, spin, and orbital periods of H115927 with a phase bin of 0.05.} 
\label{fig:tessflc_H115927}
\end{figure}

\begin{figure*}
\centering
\includegraphics[width=0.65\textwidth]{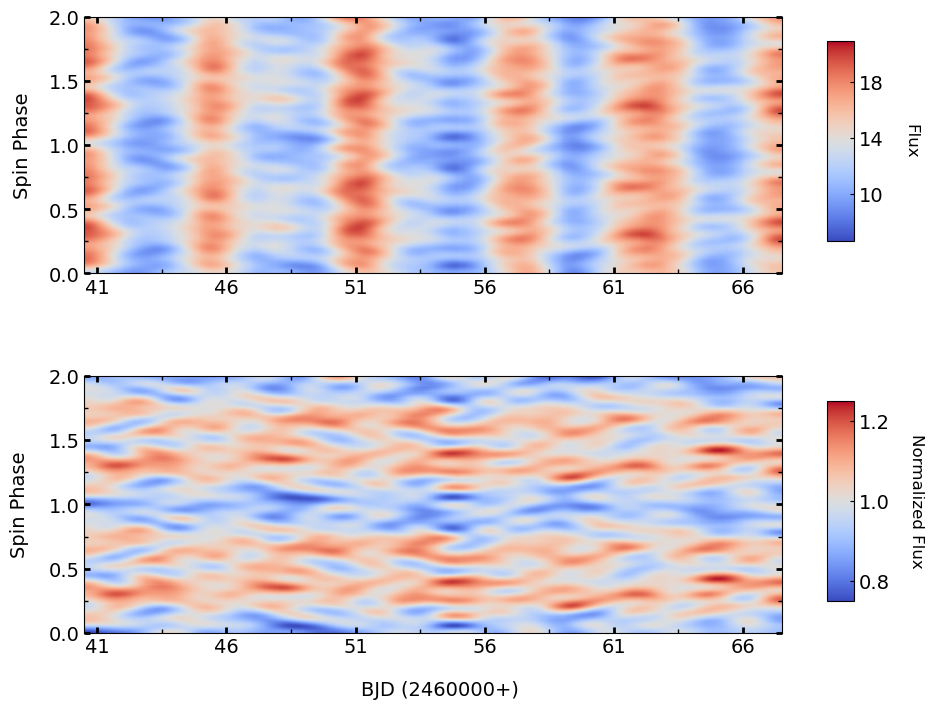}
\caption{Spin-pulse profiles of H115927 obtained from one-day TESS observations, shown for both non-normalized (top) and normalized (bottom) data, with a phase bin of 0.05.} 
\label{fig:1dayflc_H115927}
\end{figure*}

[PK2008] HalphaJ115927 (hereafter H115927)  was classified as a CV candidate due to the presence of Balmer emission lines, flat Balmer decrements, and possible He II $\lambda$4686 emission \citep{Pretorius08}. 

IGR J14091-6108 (hereafter IGRJ14091) was discovered during the INTEGRAL 9-year Galactic Hard X-ray Survey \citep{Krivonos12}. Based on Chandra observations, \cite{Tomsick16a}  classified it as most likely a magnetic CV. Later, \cite{Tomsick16b} further studied this source using X-ray and optical observations. Their optical spectra showed strong emission lines with significant variability in the lines and continuum, which indicated that they originated from an irradiated accretion disc. Moreover, based on the   X-ray data, they derived a spin period of approximately $\sim$  576.3 s, and confirmed its classification as a magnetic CV belonging to the IP class. Based on X-ray spectroscopy, they suggested that it is a candidate system for having a higher mass than a typical WD, possibly near the Chandrasekhar limit.

Past studies reveal that H115927 has been poorly studied, leaving its key properties and behavior entirely unknown. The light curve morphology and periodic behavior of this system  have not been determined yet, as it is a newly identified CV candidate. It is also unknown which specific class of CV it belongs to, if it is indeed a CV. The limited research on this object encourages a more in-depth study. On the other hand, the absence of an orbital signal in IGRJ14091 has prompted for a re-examination of its properties, as the orbital period is crucial to understanding the nature of such systems. To address these gaps, precise optical analyses using the Transiting Exoplanet Survey Satellite \citep[TESS;][]{Ricker15} data were conducted for both systems. The high-cadence, long-span, and uninterrupted nature of the TESS data provides an excellent opportunity to investigate long-term and short-term variations in both systems, which are important for understanding of CVs in general.

The paper is organized as follows: the observations and data reduction are summarized in the next section. Section \ref{sec:analysis} contains analyses and the results. Finally, the discussion and conclusions are presented in Sects. \ref{sec:disc} and \ref{sec:sum}, respectively.


\section{Observations and data reduction}\label{sec:obs}
H115927 was observed by the TESS from 2023-04-06T14:48:04.533 to 2023-05-03T12:50:04.378, with a cadence of 120 s, for a duration of 26.9 d. On the other hand, IGRJ14091 was observed with both 120 s and 20 s cadences from 2023-05-04T05:48:29.056 to 2023-06-01T03:10:25.085, for a duration of 27.8 d.  
 The data presented in this work were obtained from the Mikulski Archive for Space Telescopes (MAST) at the Space Telescope Science Institute. The specific observations analyzed can be accessed through \dataset[doi:10.17909/kevb-8226]{https://archive.stsci.edu/doi/resolve/resolve.html?doi=10.17909/kevb-8226}. The data is stored with identification numbers `TIC 938248761' and `TIC 1027495091' for H115927 and IGRJ14091, respectively. TESS consists of four cameras, each with a field of view of 24$\times$24 degree$^2$, which are aligned to cover 24$\times$96-degree$^2$ strips of the sky called `sectors' \citep[see][for details]{Ricker15}. H115927 and IGRJ14091 were observed by TESS in sectors 64 and 65, respectively. TESS bandpass extends from 600 to 1000 nm, with an effective wavelength of 800 nm. The TESS pipeline provides Simple Aperture Photometry (SAP) and Pre-search Data Conditioned SAP (PDCSAP) flux values. The PDCSAP light curve attempts to remove instrumental systematic variations by fitting and removing the signals that are common to all stars on the same CCD\footnote{\url{https://outerspace.stsci.edu/display/TESS/2.1+Levels+of+data+processing}} . By employing this approach, aperiodic variability might be removed from the data. To confirm that the PDCSAP data retain all aperiodic variability, a periodogram analysis was performed on both sources, considering both SAP and PDCSAP data. The PDCSAP and SAP light curves showed consistent short-period variability for both systems, indicating that the PDCSAP data are reliable and can be used for the further analysis. Anomalous event-related data displayed quality flags above zero in the FITS file structure.  Therefore, only the data with QUALITY flag of zero are considered, using the PDCSAP flux values.


\section{Analysis and results}\label{sec:analysis}
\subsection{[PK2008] HalphaJ115927}
 The TESS light curve of H115927, depicted in Figure \ref{fig:tesslc_H115927}, clearly exhibits a periodic pattern. To search for periodic signals, a period analysis was conducted using the Lomb-Scargle (LS) periodogram algorithm \citep[][]{Lomb76, Scargle82}. The resulting LS power spectrum is shown in Figure \ref{fig:tessps_H115927}, marks the positions of all identified frequencies. The power spectrum features a prominent peak corresponding to a period of 1161.49$\pm$0.14 s ($\sim$ 74.38 d$^{-1}$). Additionally, a strong signal is detected in the lower-frequency region at a period of 7.20$\pm$ 0.02 h ($\sim$ 3.33 d$^{-1}$). Notably, another peak at 1215.99$\pm$0.15 s ($\sim$ 71.05 d$^{-1}$) is also present, which precisely matches the expected beat period, calculated as the difference between the $\sim$ 74.38 d$^{-1}$ and 3.33 d$^{-1}$ signals. This observed beat period supports the identification of $\sim$ 1161.49 s and  $\sim$ 7.2  h as the spin ($P_\omega$) and orbital ($P_\Omega$) periods, respectively. Other periods, corresponding to harmonics of the spin period, specifically $P_{2\omega}$ and $P_{3\omega}$, are also detected in the power spectrum and are given in Table \ref{tab:ps}. Moreover, a significant peak corresponding to a period of 3.702$\pm$0.005 h is observed after the $\Omega$ frequency (indicated by an arrow). However, this slightly deviates from  the expected second harmonic of the orbital period ($P_{2\Omega}$) of 3.6 h. Additionally, a strong signal corresponding to a period of 5.66$\pm$0.29 d is present in the low-frequency region of the power spectrum, as shown clearly in the inset panel of Figure \ref{fig:tessps_H115927}. The light curve clearly exhibits a long-term modulation in the system's brightness, corresponding to this period. The modulation at this period is also preserved in the SAP data and is therefore probably real. It seems that this might be the result of a beat between orbital variations and another periodicity. Although, no other significant signal, corresponding to any other periodicity or superhumps, is detected in the power spectrum obtained from the current observations. Using the extensive TESS observations, the temporal evolution of the power spectrum was also computed in successive one-day intervals. Employing LS periodogram analysis on each one-day dataset, a trailed power spectrum with one-day increments was constructed and shown in Figure \ref{fig:tessps_1day_H115927}. The spin period is prominent in most segments, although its amplitude fluctuates and is sometimes weakly detected. In contrast, the orbital and beat frequencies are generally weak but become noticeable in a few segments or instances.

\begin{figure*}
\centering
\includegraphics[width=0.7\textwidth]{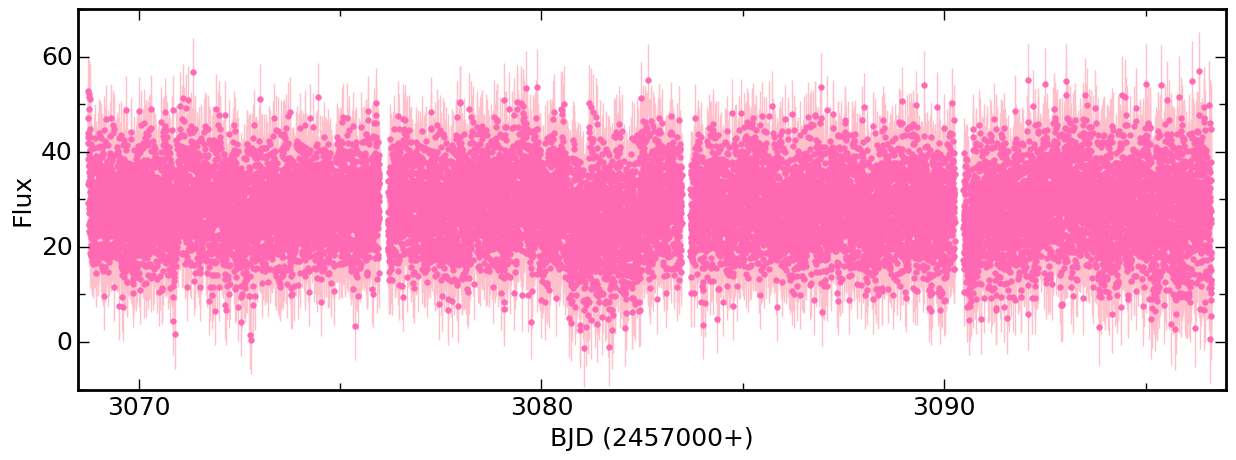}
\caption{TESS light curve of IGRJ14091.} 
\label{fig:tesslc_IGRJ14091}
\end{figure*}
\begin{figure*}
\centering
\includegraphics[width=0.7\textwidth]{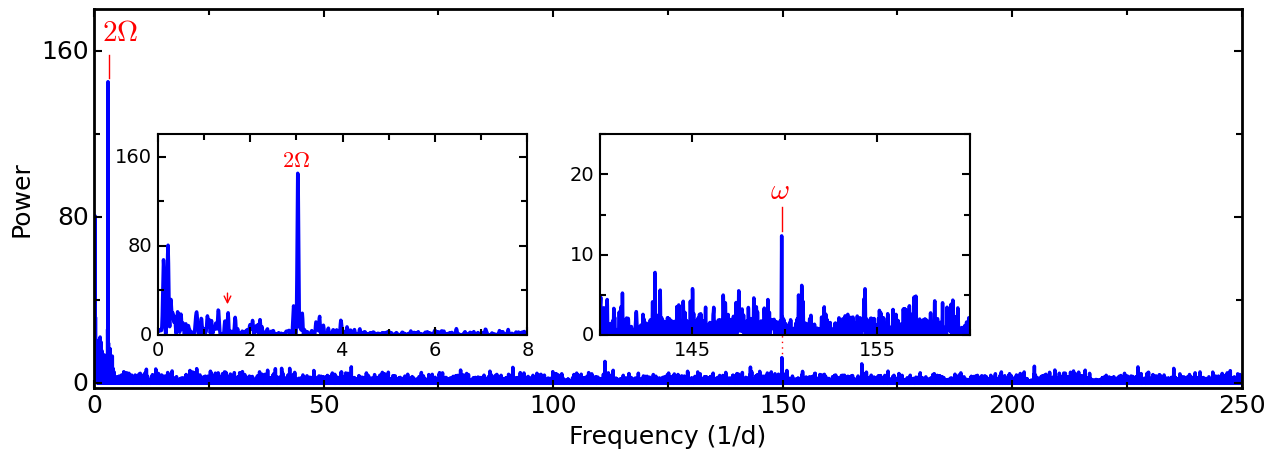}
\caption{TESS power spectrum of IGRJ14091 obtained from 120 s cadence TESS data, with inset panels representing zoomed-in plots of the $\omega$ and $\Omega$ frequency regions. The significant frequencies observed in the power spectrum are marked.} 
\label{fig:tessps_IGRJ14091}
\end{figure*}

The periodic variability in H115927 was further investigated by folding the light curves over derived 
 orbital, spin, beat, and longest periods, using an arbitrary zero time BJD = 2460041.1182 (chosen first point of the TESS observation) as the reference epoch. The phase-folded light curves were generated using the combined TESS data, with a binning of 20 points per phase, and are presented in Figure \ref{fig:tessflc_H115927}. The orbital- and beat-phased light curves exhibit noticeable periodic modulations with a single maximum. However, the spin-phased light curve displays non-sinusoidal and appears to be double-peaked modulations with the maxima occurring near phases 0.35 and 0.65, separated by a shallow minimum near phase 0.45. A strong, broad humped modulation splitting into two separate peaks, along with a shallow dip near phase 0.94, is also observed corresponding to the longest period of $\sim$ 5.66 d. To inspect the evolution of pulse profiles, the day-to-day periodic spin variations were also analyzed. Following the same approach as mentioned above, each day's light curve was folded using the derived spin period with a binning of 20 points in a phase. Figure \ref{fig:1dayflc_H115927} represents the color-composite plot for the one-day spin-phased pulse profile for H115927. Similar to the combined light curve, double-peaked spin modulations with a minimum at 1.0 are observed throughout the one-day TESS observations, which can be clearly seen in the normalized light curves (see bottom panel of Figure \ref{fig:1dayflc_H115927}). However, the amplitude of spin pulse profile is highly variable. For approximately three days, the amplitude is higher, followed by a decrease over the next three days. This recurring cycle is observed across the 1-day segments and is distinctly visible in the non-normalized light curves (see top panel of Figure \ref{fig:1dayflc_H115927}).

\subsection{IGR J14091-6108}
\begin{figure}
\centering
\includegraphics[width=0.44\textwidth]{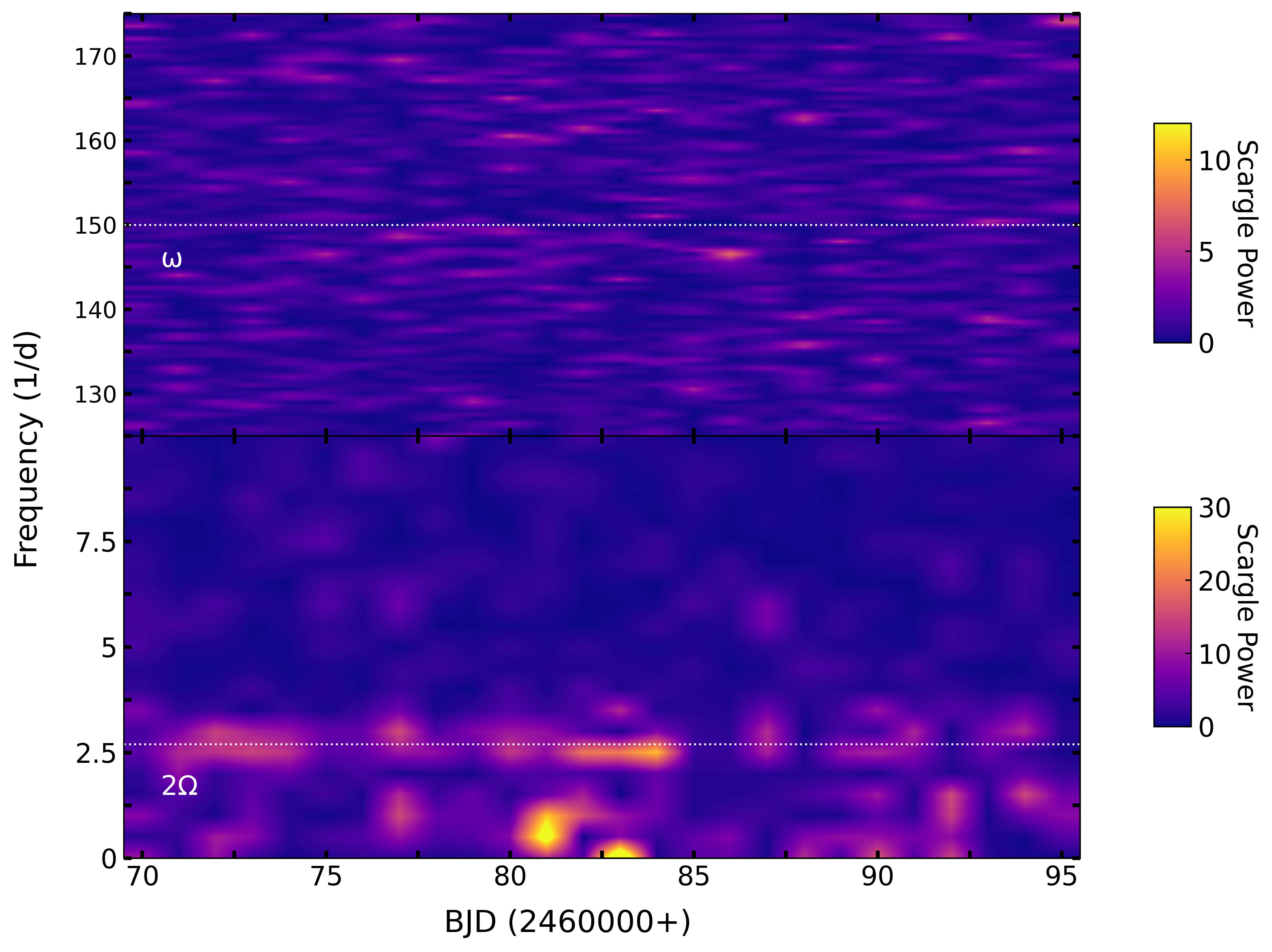}
\caption{Time-resolved power spectra of IGRJ14091 computed using a 1-day wide moving window across the entire dataset of the TESS observations near $2\Omega$ and $\omega$ frequency regions.} 
\label{fig:tess_1dayps_IGRJ14091}
\end{figure}

\begin{figure}
\centering
\includegraphics[width=0.45\textwidth]{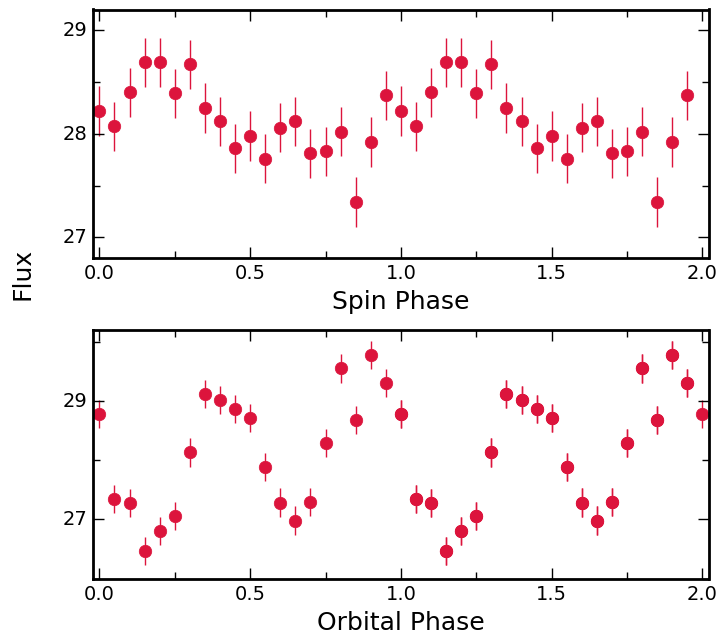}
\caption{Orbital-and-spin phase folded light curves of IGRJ14091 with a phase bin of 0.05.} 
\label{fig:tessflc_IGRJ14091}
\end{figure}

TESS light curve of IGRJ14091 is presented in Figure \ref{fig:tesslc_IGRJ14091}, revealing short-term fluctuations combined with long-term variability upon closer inspection. To identify the periodicities, the LS periodogram analysis was performed on the entire TESS dataset for cadence 120 s, as shown in Figure \ref{fig:tessps_IGRJ14091}, where the positions of the identified frequencies are marked. The detected significant frequency at $\sim$ 150 d$^{-1}$ corresponds to the spin period of 576.63$\pm$0.03 (highlighted in the right-side inset panel), confirming the previous findings of \cite{Tomsick16b}. Another prominent peak is observed in the lower frequency region at $\sim$ 3 d$^{-1}$ in the power spectrum. This strong peak corresponds to a period of 7.92$\pm$0.02 h. Folding of the light curve at twice of this period (15.84 h) reveals two distinct maxima and minima (see bottom panel of Figure \ref{fig:tessflc_IGRJ14091}), implying the possibility that the 7.92 h period is a second harmonic of the 15.84 h orbital period. This 15.84 h signal is visible in the power spectrum and is marked in the left-hand inset panel, although it appears with relatively low power. This inferred signal as the orbital period is further discussed in detail below by the distinctive double-peaked feature observed in its folded light curve. Similar to H115927, a one-day time-resolved power spectrum analysis was conducted for IGRJ14091. The spin frequency was not found to be significant in the power spectrum of each day segment. However, the harmonic of the orbital frequency showed irregularities throughout the time-resolved power spectrum, with significant power most of the days and relatively small power in a few instances (see Figure \ref{fig:tess_1dayps_IGRJ14091}).

\begin{figure}
\centering
\includegraphics[width=0.5\textwidth]{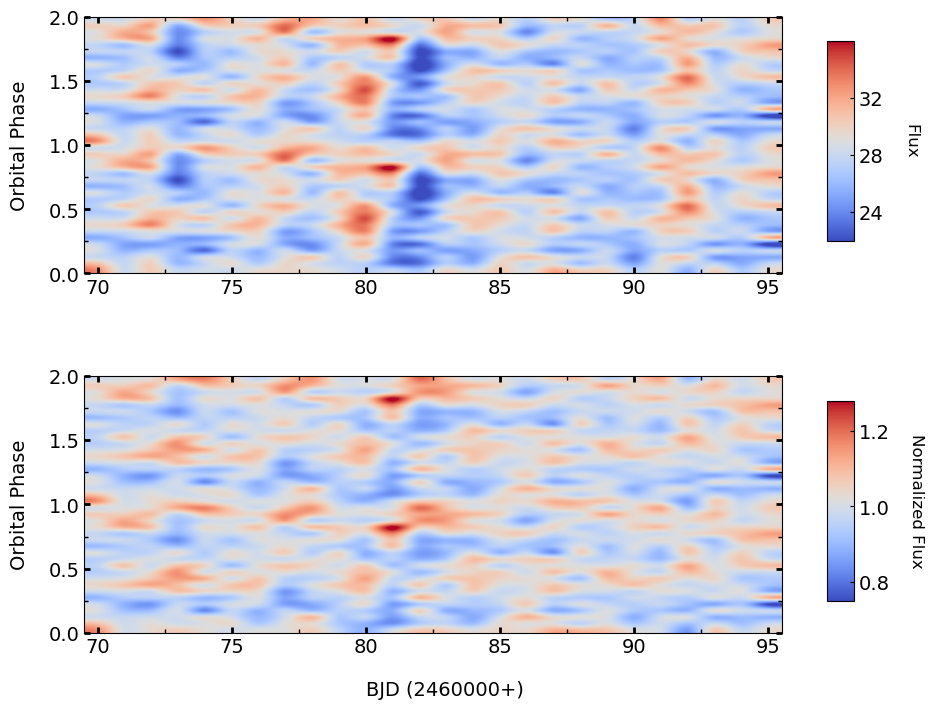}
\caption{Orbital-pulse profiles of IGRJ14091 obtained from one-day TESS observations, shown for both non-normalized (top) and normalized (bottom) data, with a phase bin of 0.05.} 
\label{fig:tess_1dayflc_IGRJ14091}
\end{figure}

To examine the phased-light curve variations of IGRJ14091, the entire TESS dataset was folded using the first point of the TESS observation as the reference epoch, BJD=2460068.7434, and the derived spin and orbital periods. The folded light curves with a binning of 20 points per phase are shown in Figure \ref{fig:tessflc_IGRJ14091}. The spin-phased light curve exhibits modulations resembling a sinusoidal pattern, with a maximum near phase 0.2.  However, the orbital-phased light curve reveals a double-peaked profile, with maxima near phases 0.4 and 0.9. The difference between these two peaks in the folded light curve is statistically significant at the 2.2$\sigma$ level. This measurable asymmetry further suggests that the fundamental period is 15.84 h, which corresponds to the orbital period of the system.  
 Additionally, both a single-sinusoidal and a double-sinusoidal function (each with a constant term) were fitted to the folded light curve. The single-sinusoidal function failed to fit the double-peaked structure, while the double-sinusoidal function provided a significantly better fit, closely matched the observed profile. This two-component sinusoidal fit reflects that there are indeed two distinct peaks in the folded light curve and provides additional confirmation for the orbital period of the system. One-day phased light curve variations during orbital motion were also constructed for IGRJ14091. Similar double-peaked features with variable amplitudes are observed in one-day segments of the orbital-phased light curves, as depicted in Figure   \ref{fig:tess_1dayflc_IGRJ14091}.

\section{Discussion}
\label{sec:disc}
  Detailed time-resolved analyses were carried out for H115927 and IGRJ14091, using high-cadence optical photometric data from the TESS. Considering $\sim$ 1161.5 s and $\sim$ 7.2 h as spin and orbital periods, respectively, H115927 is speculated to belong to the IP class of magnetic CVs. Whereas the detection of a spin signal of $\sim$ 576.6 s and previously unknown probable orbital signal of $\sim$ 15.84 h in IGRJ14091, supports the classification of this system as an IP.
  
  For H115927, six significant periods were observed, including two fundamental periods at  7.20$\pm$0.02 h and 1161.49$\pm$0.14 s, which can be interpreted as the probable orbital and spin periods, respectively. Considering these spin and orbital periods, the beat period is estimated to be $\sim$ 1216 s, which is also present in the power spectrum at period 1215.99$\pm$0.15 s. These periods are newly identified and reported here for the first time. The presence of these multiple frequencies indicates that H115927 likely belongs to the IP class of magnetic CVs. The possible identification as an IP could explain the detection of moderate He II $\lambda$4686 emission in the optical spectrum of H115927, as presented by \cite{Pretorius08}. The presence of this emission line suggests high-energy photons, a characteristic feature typically associated with magnetic CVs, high mass-transfer nova-like variables, and post-novae.  H115927 also exhibits strong periodic variations on a timescale of 5.66$\pm$0.29 d, which likely correspond to the precession period of the accretion disc. This observed long-period modulation makes H115927 a particularly intriguing object, similar to the IP TV Col, which also exhibits long-period variations on a timescale of $\sim$ 4 days \citep[see][]{Motch81, Hutchings81, Barrett88, Bruch22}. A strong spin pulse and its harmonics (2$\omega$ and 3$\omega$) dominate the power spectrum of H115927, whereas $\Omega$ and $\omega-\Omega$ are detected with comparatively lower amplitudes. In IPs, the emergence of the combined spin and orbital frequencies as $\omega - \Omega$ in optical is often seen and is likely caused by the reprocessing of X-rays in structures fixed in the binary frame, such as the secondary and/or the bright spot region of the accretion disc \citep[][]{Hassall81, Patterson81}. Thus, X-ray reprocessing by the hot spot region of the accretion disc or the secondary is likely responsible for the optical modulation at the beat period in H115927. In contrast to optical, the detection of a beat pulse in X-rays characterizes the mode of accretion in IPs. The presence of both $\omega$ and $\omega - \Omega$ in X-rays typically suggests disc-overflow accretion \citep[see][]{Hellier93}. Since this source has not been observed in X-rays so far, we do not know if there is X-ray beat modulation. It remains a possibility that beat modulation  could be identified in future X-ray observations, which may suggest that this system is a disc-overflow. For now, it is presumed to be a disc-fed accretor. 

\begin{figure}
\centering
\includegraphics[width=0.48\textwidth]{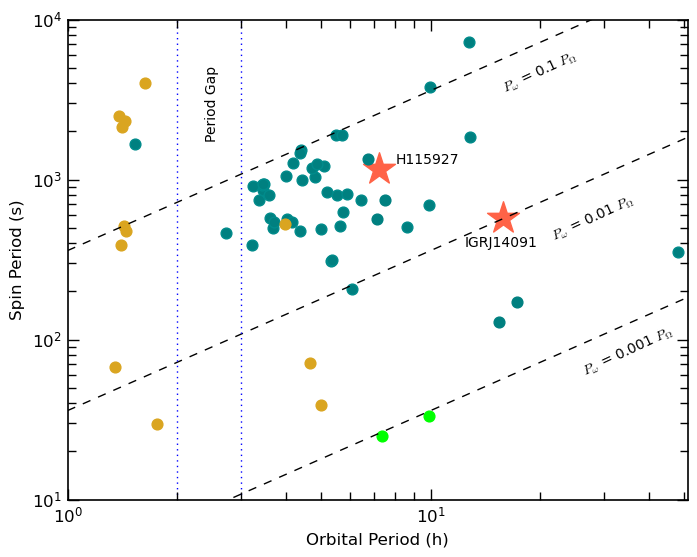}
\caption{Spin-and-orbital period distribution of the confirmed IPs compiled by Mukai's catalog. Light green symbols denote propeller systems, golden circles represent low-luminosity IPs. The symbol star represents H115927 and IGRJ14091 from the present study. } 
\label{fig:dis_IPs}
\end{figure}

For IGRJ14091 a significant period of 576.63$\pm$0.03 s is detected, which corresponds to the spin period of the system and is well consistent with the period reported by \cite{Tomsick16b}. Due to the longer-baseline of TESS, a strong significant period of 7.92$\pm$0.02 h is also observed in its power spectrum. This detected period appears to be  a harmonic of the system's orbital period, implying an orbital period of $\sim$ 15.84 h. This period has not been detected in earlier studies, as \cite{Tomsick16b} did not search for the orbital period. The dominant peak at $2\Omega$ and the observed double-humped features in the orbital-phase light curve  suggest that these variations result from ellipsoidal modulation of the secondary \citep{Warner95}. The existence of both spin and orbital periods supports its classification as an IP. Besides the orbital and spin periods, no other sideband frequencies are present in its power spectrum. Interestingly, very few systems have been found that belong to the category of long-period confirmed IPs, such as RX J2015.6+3711 \citep[$P_\Omega$=12.7 h;][]{Halpern18}, GK Per \citep[$P_\Omega$=47.9 h;][]{Crampton86}, Swift J1701.3$-$4304 \citep[$P_\Omega$=12.8 h;][]{Shara17}, V2731 Oph \citep[$P_\Omega$=15.4 h;][]{Gansicke05}, and Swift J2006.4+3645\citep[$P_\Omega$=17.3 h;][]{Hare21}. The inferred orbital period of 15.84 h in IGRJ14091 adds to the population of long-period IPs, providing a new example for studying such systems.  With the the detected spin period of $\sim$ 576.6 s and $P_\omega$/$P_\Omega$ $\sim$ 0.01, this system appears to belongs to the fast rotator group of disc-fed IPs. 

The observed spin modulations in H115927 are non-sinusoidal compared to most IPs and exhibit a double-peaked pulse profile, as expected from the observed second  harmonic of the spin pulse in its power spectrum. Although, the amplitude of the second harmonic 2$\omega$ is weaker than the fundamental frequency $\omega$. This observed double-humped feature can be interpreted in terms of two emitting poles. Both humps exhibit comparable amplitudes, with peaks separated by approximately 0.3 in phase, indicating that the two poles may accrete at similar rates and are separated by approximately 110$^\circ$. This suggests that both poles contribute significantly to the spin modulation. Although the accreting poles are not exactly opposite, their geometry allows for equal visibility of both accreting poles, producing two distinct maxima. On the other hand, a single-peaked sinusoidal like spin pulse profile observed in IGRJ14091, commonly seen in many IPs, can be attributed to the changing visibility of the accretion curtains caused by a relatively low dipole inclination. This results in a single-peaked pulse, with the maximum occurring when the upper pole points away from observer \citep{Hellier91, Kim96}.  

Using Mukai's catalog of IPs\footnote{\url{https://asd.gsfc.nasa.gov/Koji.Mukai/iphome/iphome.html}}, an updated spin-orbital period diagram has been constructed that includes H115927 and IGRJ14091 along with other confirmed IPs (see Figure \ref{fig:dis_IPs}). The observed orbital periods for both systems place them above the period gap, with spin-to-orbital period ratios ($P_\omega$/$P_\Omega$) of $\sim$ 0.045 and $\sim$ 0.01 for H115927 and IGRJ14091, respectively. Using the \cite{Patterson94} relation $P_{\text{spin}} - \log{\mu}$, the magnetic moment of the WD ($\mu$) for H115927 and IGRJ14091 can be estimated as $2.2\times10^{33}$ G cm$^{3}$ and $1\times10^{33}$ G cm$^{3}$, respectively. \cite{Norton04} illustrated a broad range of spin equilibria exists in the parameter space of $P_\omega$/$P_\Omega$, $P_\Omega$, $\mu$, and $q$ (mass ratio), as shown for $q$ of 0.5 in their figure 2. Assuming that the majority of the observed IPs are close to their spin equilibria, \cite{Norton04} estimated their $\mu$ from the model calculations. Using their figure 2, the magnetic moment of the WD can be estimated as $\mu$ $\sim$ 1 $-$ 2 $\times$ 10$^{33}$ G cm$^{3}$ for
  $P_\Omega$ $\sim$ 7.2 h and 
 $P_\omega$/$P_\Omega$ $\sim$ 0.045 of H115927, which closely aligns with the result from \cite{Patterson94}. Although, for an accurate determination of the magnetic moment, it is essential to know the mass ratio and the accretion rate of both  systems.  Using the derived orbital period of H115927, the mean density of the secondary is estimated to be approximately 2.1 g cm$^{-3}$  \citep[see][]{Warner95}, which falls in the range of mean densities of a main-sequence star of G-K type \citep{Allen76}. However, for the long orbital period of IGRJ14091, the estimated density of the secondary is 0.4 g cm$^{-3}$, which is too low for it to be a main-sequence star, suggesting that it may have an evolved donor, similar to V2731 Oph.


\section{Conclusions}\label{sec:sum}
 Detailed analyses of a candidate CV H115927 is presented using optical TESS photometry. For the first time, this study reports periods of $\sim$ 7.2 h, $\sim$ 1161.5 s, and $\sim$ 1215.9 s  for this system, which are considered the  probable orbital, spin, and beat periods of the system, respectively. The detection of both long period and short-timescale variabilities suggest that this system likely belongs to the IP class of magnetic CVs. Interestingly, H115927 also exhibits strong periodic variations on a timescale of $\sim$ 5.6  d, suggesting that this may result from the precession of an accretion disc, similar to the IP TV Col. The strong $\sim$ 7.92 h period observed in IGRJ14091 can be interpreted as the second harmonic of the $\sim$ 15.84 h orbital period. The spin signal of $\sim$ 576.6 s and orbital signal of $\sim$ 15.84 h supports the classification of this system as an IP. The inferred orbital period of $\sim$ 15.84 h in IGRJ14091 adds a new sample to the limited population of long-period IPs. The observed dominant signal at the second harmonic of the orbital frequency and double-humped orbital modulation in IGRJ14091 are suggestive of ellipsoidal modulation of the secondary. The observed strong double-peaked optical spin pulse profile in H115927, with equal maxima and unequal minima, could result from two-pole accretion. Both poles contribute to the spin modulation, and their geometry ensures equal visibility of both accreting poles, producing distinct maxima.  In contrast, IGR J14091-610 exhibits a single-peaked sinusoidal spin pulse, attributed to the changing visibility of the accretion curtains due to a relatively low dipole inclination. The present observations imply that both systems are likely disc-fed accretors. 



\section*{Acknowledgments}
I thank the anonymous referee for providing useful comments and suggestions that led to the significant improvement of the quality of the paper. AJ acknowledges support from the Centro de Astrofisica y Tecnologias Afines (CATA) fellowship via grant Agencia Nacional de Investigacion y Desarrollo (ANID), BASAL FB210003. This paper includes data collected with the \textit {TESS} mission, obtained from the MAST data archive at the Space Telescope Science Institute (STScI). Funding for the \textit {TESS} mission is provided by the NASA Explorer Program.


\bibliography{ref}{}
\bibliographystyle{aasjournal}

\end{document}